\documentclass[a4paper,fleqn]{cas-sc}

\usepackage[numbers]{natbib}
\usepackage{graphicx}
\usepackage{todonotes}

\usepackage{tabularx}
\usepackage{array}
\usepackage{booktabs}

\def\tsc#1{\csdef{#1}{\textsc{\lowercase{#1}}\xspace}}
\tsc{WGM}
\tsc{QE}

\begin{document}

\let\WriteBookmarks\relax
\def\floatpagepagefraction{1}
\def\textpagefraction{.001}

% =========================================================
% Title
% =========================================================

\shorttitle{Malware Clustering in the Wild}

\shortauthors{Raguvir S et al.}

\title[mode=title]
{Malware Clustering in the Wild: Taxonomy, Challenges, and Evaluation Metrics}

% =========================================================
% Authors
% =========================================================

\author[1]{Raguvir S}
\cormark[1]
\ead{nv25812@umbc.edu}

\author[1]{Bojing Li}
\ead{ji18978@umbc.edu}

\author[1]{Prajna Bhandary}
\ead{prajnab1@umbc.edu}

\author[1]{Charles Nicholas}
\ead{nicholas@umbc.edu}

% =========================================================
% Corresponding author
% =========================================================

\cortext[1]{Corresponding author}

% =========================================================
% Affiliation
% =========================================================

\affiliation[1]{
    organization={University of Maryland, Baltimore County},
    addressline={Department of Computer Science and Electrical Engineering},
    city={Baltimore},
    postcode={21250},
    state={Maryland},
    country={USA}
}

% =========================================================
% Abstract
% =========================================================

\begin{abstract}
As malware continues to grow in volume, diversity, and sophistication, clustering has emerged as a critical technique for analyzing large malware corpora. 
Clustering enables the discovery of structural relationships, behavioral patterns, and evolutionary trends, and is widely used to group unknown samples, infer malware family boundaries, and identify novel variants. 
However, existing research remains fragmented, employing a wide range of clustering techniques with no unified perspective that systematically organizes prior work or clearly identifies gaps in the literature.
This survey addresses this gap by organizing previous malware clustering research into seven categories: centroid-based, hierarchical-based, density-based, distribution-based, mixed-type, fuzzy, and AI-driven approaches. 
In addition, this survey summarizes publicly available malware datasets commonly used in clustering research and examines the feature representations derived from these datasets. 
Our goal is to clarify the current landscape and outline promising avenues for future research on malware clustering, including how AI-driven methods may shape large-scale and reproducible analysis.
\end{abstract}

% =========================================================
% Keywords
% =========================================================

\begin{keywords}
Malware clustering
\sep Malware analysis
\sep Malware classification
\sep Clustering
\sep Cybersecurity
\end{keywords}

% =========================================================
% Create title
% =========================================================

\maketitle

% =========================================================
% Main text
% =========================================================

\section{Introduction}

% With growing recognition of machine learning (ML) methods and their applications in various fields, it was inevitable for the cybersecurity space to adopt these techniques. Numerous research has explored how ML algorithms can be used and tuned for the purposes such as malware detection and identification. 

Malware continues to evolve in scale, sophistication, and strategic intent, challenging defenders to recognize and respond to emerging threats at operational speed. Although signature-based detection remains foundational, the rapid growth of polymorphism, obfuscation, rebranding, and automated malware generation has shifted the community toward behavior- and structure-driven approaches. Within this landscape, clustering has emerged as a powerful analytical method. Unlike traditional classification methods, clustering does not assume ground truth labels; instead, it groups samples based on inherent structural, behavioral, or semantic similarity. These clusters often reveal familial relationships, campaign linkages, shared code patterns, and evolutionary trajectories, enabling analysts to better understand malware behavior through its underlying features.

Despite its long-standing use in malware analysis research, relatively few studies systematically examine the breadth of clustering methodologies applied in this domain. Much is known about how malware families can be grouped using static features, dynamic features, or hybrid representations, but important questions remain unanswered: Are there areas of malware analysis in which the use of clustering has yet to be explored? Where do we still lack empirical evaluation, methodological rigor, or dataset diversity? As adversaries expand the scale and complexity of malware campaigns, understanding the strengths and limitations of existing clustering strategies becomes essential for both research and practice.

%% perhaps writing this would be a repeat
% This survey addresses these questions by providing a structured overview of the major clustering paradigms applied to malware:
% (i) Centroid-based 

A second motivation for this survey is the lack of a unified view of the datasets used in malware clustering research. Although many studies rely on publicly available corpora, their use is inconsistent: some focus on small or outdated collections, others depend on restricted or private datasets, and very few consider the impact of temporal drift, label noise, or feature imbalance. Because clustering outcomes are deeply impacted by data distribution, diversity of families, and the richness of available features, these inconsistencies make it difficult to compare findings across studies or assess how well different methods generalize. By consolidating publicly documented malware datasets released after 2020, this survey provides an up-to-date overview of available data resources and highlights empirical gaps that limit reproducibility and large-scale evaluation.

%% Our contribution

This survey addresses these challenges by providing a structured overview of malware clustering research across seven methodological categories, spanning classical and AI-driven approaches. In addition, we summarize publicly available datasets used in prior studies and discuss how current practices shape empirical evaluation. Together, these perspectives aim to clarify the current research landscape and to identify promising directions for advancing malware clustering in the face of evolving threats.

%%Note: we must also include the feature aspect of it

%% Why is malware important?

%% Why malware clustering?

%% challenges in malware clustering? or scope?

\section{Related Work}

% \todo[inline]{Draft ready- Related Work}
Previous work has explored the applications of machine learning in malware analysis and the challenges it brings. Although these studies provide important foundations, clustering is often discussed as one component of a broader malware-analysis pipeline rather than as a research area requiring its own systematic synthesis.

This section reviews prior studies on malware detection and identification, focusing specifically on the application of clustering algorithms. First we discuss previous literature surveys that review the applications of machine learning to detect and identify malware, covering features, methodologies and datasets used. Following this, we explore the limitations observed in these studies and the state of this area of study. 

%\gripe{Before we launch into discussing papers} let's explain how we organized this section.  Do we discuss older papers first?  Broader surveys?  And it's important to decide and describe the criteria we use for assessing the papers.  At this point, the presentation seems random.
%\gripe{help the reader} understand the structure AND the narrative flow of the paper.  No paper should simply be described in a sentence or two within establishing a foundation and a context for that point in the discussion.
We begin with surveys that cover machine learning for malware analysis broadly.
Ucci et al.~\cite{ucci2019survey} provide a comprehensive collection of research topics in machine learning for malware analysis including detection, similarity analysis, category detection, feature extraction, supervised, unsupervised, and semi-supervised machine learning algorithms used for malware categorization. The study also identifies several challenges associated with applying machine learning to malware data. Ye et al.~\cite{ye2017survey} discusses approaches to detecting and identifying similarities between malware, explaining how malware features, such as API calls, strings, opcodes, and control flow graphs are used in machine learning techniques.
Although both surveys established a useful overview of the field, their broad scope limits its treatment of clustering-specific considerations, such as algorithm selection and cluster validation. %This survey provides a detailed overview of recent research on malware analysis.

We now turn to work centered on clustering, Bayer et al.~\cite{bayer2009scalable} extract fine-grained dynamic behavior as precise features and apply the combination of Jaccard similarity and locality-sensitive hashing for large-scale clustering. Compared to compression-based approaches, the framework significantly enhances scalability while maintaining high clustering quality. This work provides a comprehensive study of various features of malware, including system call features, network logs, and control flow graphs. The study further explores how these features can be used to create a behavioral profile of malware, which can in turn be used in clustering. Given the in depth discussion of malware features and clustering algorithms, this study provides much-needed groundwork for numerous future malware behavior-related studies. Amira et al.~\cite{amira2023survey} introduced a survey on community detection in malware analysis.
By constructing a similarity network of malware specimens based on feature similarity and applying community-detection algorithms, structurally clustered communities can be identified.
% explores the concept of clustering malware based on how they behave, this work shows that 
%\gripe{add to this?}
 
%% Why is our work different
Taken together, these works provide a good baseline for understanding malware clustering, to our knowledge, there is no systematic review that comprehensively summarizes and distills the latest advancements in this field. On one hand, there is no comparative analysis of mainstream clustering methods for malware analysis. On the other hand, there is a lack of summaries regarding the applicability of different clustering methods across various malware analysis tasks. 
Moreover, guidelines for selecting clustering methods according to the characteristics of different malware features are also needed. Furthermore, currently available and representative malware datasets are often described and presented separately. 
This paper aims to provide a systematic synthesis of the fragmented information above, offering a structured reference for future malware analysis research and practical applications.

\section{Characterization of survey papers based on Clustering Approaches}

% \gripe{maybe this explanation needs to go into the previous section}
To organize the landscape of malware clustering research, we group existing work according to the underlying clustering paradigm. Although many studies differ in feature representation, evaluation strategy, or application domain, the core algorithms they employ fall into a small number of methodological families. Examining clustering techniques through these categories allows us to compare how each approach behaves under different data conditions, what assumptions it makes about malware structure, and where it has proven effective or limited. The following subsections outline these major clustering families, highlighting their principles, their use in malware analysis, and the unique challenges they introduce.

\subsection{Centroid-Based Techniques}
\label{sec:centroid}
%%General idea of the technique

Centroid-based clustering has played a central role in malware clustering research due to its scalability and conceptual simplicity. Across the literature, however, we find that the \emph{K-means} algorithm is used more often than any other clustering algorithm. Despite the availability of other centroid-based techniques (e.g., $k$-medoids, Fuzzy C-means, Mini batch K-means, K-Means++), most studies continue to rely on K-means or its close variants, as summarized in Table~\ref{tab:centroid-based}.  

%%relationship between families
% \gripe{why discuss this paper at this particular point?}
Basole and Stamp~\cite{basole_cluster_2021} examine 20{,}000 malware samples across 20 families from the Malicia corpus and study alignment between cluster structure and family labels, showing that centroid partitions can capture taxonomic structure when supported by sensible preprocessing and feature selection.

At larger scales, centroid-based clustering is increasingly used as a structural or organizational step rather than a standalone analytical tool. Jure\v{c}kov'a et al. ~\cite{jureckova_classification_2024} leverage hundreds of thousands of PE samples from the EMBER dataset, where clustering serves primarily as a preprocessing step within a larger hybrid pipeline that combines unsupervised grouping with supervised refinement. This approach reflects a broader trend toward using fast centroid-based methods to impose coarse structure on malware corpora before downstream learning or classification.

Across these studies, the dominant factor influencing clustering performance is not the specific centroid-based algorithm, but the geometry implied by the features and the corresponding distance metric:
\begin{itemize}
\item \textbf{High-dimensional sparse static features} such as strings or $n$-grams benefit from cosine similarity or spherical K-means, or from careful normalization such as TF--IDF 
% \gripe{this sentence doesn't yet make sense}
prior to Euclidean K-means, which mitigates length effects and improves cluster compactness \cite{pai_clustering_2017}.

\item \textbf{Behavioral counts} including registry, API, or event frequencies perform well with Euclidean K-means when counts are appropriately scaled using log transforms or TF--IDF weighting, often yielding larger gains than changing the clustering algorithm itself \cite{rosli_clustering_2019}.

\item \textbf{Dense or correlated numeric vectors} use whitening or a Mahalanobis-based centroid 
%\todo{check use of M-based vs. M-aware}
criterion, or apply feature weighting/rescaling (as in \cite{cordeiro_de_amorim_identifying_2021, duraisamy_soundrapandian_mulber_2022}) to reduce overlap and sharpen separation.  (By Mahalanobis-based, we refer to use of a distance function that takes covariance into account, that is, how features vary together.)

\item \textbf{Large-scale or streaming settings} commonly employ mini-batch or online K-means to quantize structure, then feed cluster indices/centroids to maintain efficiency, with periodic monitoring using internal validity metrics such as Silhouette score ~\cite{rousseeuw1987silhouettes} or Davies–Bouldin index to trigger re-centering or retraining when quality degrades~\cite{pitolli_malfamaware_2021, jureckova_classification_2024, mishra_cluster_2025}.
\end{itemize}

% \paragraph{Synthesis.}
Taken together, these studies indicate a shift away from naïve, standalone applications of K-means toward pipelines in which preprocessing choices and feature geometry play a central role. Across domains, clustering quality is shaped less by the algorithm itself than by how feature distributions are normalized, scaled, and embedded prior to centroid assignment.

Empirical evidence across the literature suggests several recurring pairings between feature types and centroid-based methods:

(i) normalize or rescale to match the feature distribution; (ii) choose a distance that respects sparsity or correlation; and (iii) integrate clusters with downstream learners at scale. At the same time, the literature still tilts heavily toward a single algorithm family. Broadening the toolkit, e.g., testing spherical K-means for lexical features, $k$-medoids when outliers are plentiful, Mahalanobis-based centroid clustering for correlated views, and online/incremental K-means under drift, would better align method choice with the statistical properties of malware data observed in practice.

Recent studies show another approach to centroid-based analysis using fuzzy clustering, which demonstrates an alternative approach to K-means.

% \begin{itemize}
%     \item For dense or correlated numeric features (selected statics, Android vectors), use whitening or Mahalanobis-based centroid clustering (or careful feature selection) to reduce cluster overlap \cite{duraisamy_soundrapandian_mulber_2022}.
%     \item For behavioral counts(registry/API/event frequencies), Euclidean 
%     k-means on scaled counts is a reliable baseline, with preprocessing (log/TF–IDF) doing the heavy lifting \cite{rosli_clustering_2019, amorim_identifying_2021}. 
%     \item For multi-view Android data, fuse or embed views first, then apply centroid clustering in the fused space so one view does not dominate \cite{chen_android_2024, padmavathi_performance_2022}.
%     \item For streams/drift, use online centroid updates with periodic health checks and re-initialization when validity metrics slide \cite{pitolli_malfamaware_2021, jureckova_classification_2024, mishra_cluster_2025}.

% \end{itemize}

% Fuuzy clustering is included as a subsection of cnetroid based since they are similar (also there is not enough papers)
\subsubsection{Fuzzy clustering}
Fuzzy clustering methods like Fuzzy C-means offer a way to cluster data into different groups without using hard set clusters. While following the idea of K-means, these algorithms use membership values, fuzzy clustering shows the probabilities of different data points being in different clusters. Fuzzy clustering has given good results for Android malware detection when used with LightGBM.  For example, Alaeiyan et al. \cite{alaeiyan_multilabel_2020} shows how fuzzy clustering can be used in the context of malware with a multilabel relevance system.

%Ransomware detection has benefited from fuzzy clustering as shown in Naik et al. \cite{naik_lockout-tagout_2019},
Naik's work \cite{naik_cyberthreat_2019}
demonstrates how fuzzy c-means can be used to track ransomware threat actors. This work shows a comparative study of K-means and Fuzzy C-means clustering, showing Fuzzy C-means performs slightly better. % could include more of Naik papers to show what work they did

The malware identification system described in \cite{haddadpajouh_mvfcc_2020} demonstrates how fuzzy clustering can be used to attribute malware payloads to various threat actors. %While the authors have used a small collection of malware samples for their work, it would be interesting to explore how this method would hold up on a larger collection. 
Although fuzzy clustering methods have commonly been used to identify malware families and in certain cases even attribute the malware to a threat actor,  little work seems to use fuzzy clustering to detect malware. Among the few examples we've seen, Taha \cite{taha_hybrid_2021} shows how a combination of fuzzy clustering and gradient boosting can be used to identify Android malware.

Together, these patterns highlight both the flexibility and the limitations of centroid-based clustering, underscoring the need for broader exploration of alternative centroid formulations that better align with the statistical properties of modern malware data. In particular, the reliance on fixed centroids and simple distance metrics motivates approaches that explicitly model data distributions and uncertainty. This shift gives rise to distribution-based clustering methods, which represent malware groups as probabilistic generative processes rather than geometric partitions.

%% note: this has to remain as htbp otherwise the tables will get misplaced

\begin{table*}[b]
\centering
\renewcommand{\arraystretch}{1.5} %line spacing
\setlength{\tabcolsep}{4pt} 

\begin{tabularx}{\textwidth}{@{} p{1.2cm} p{1.2cm} p{3cm} p{3cm} p{1.2cm} p{3.5cm} L @{}}
\toprule
Paper & Year & Feature & Dataset & Size & Model & Metrics \\
\midrule

\cite{basole_cluster_2021} & 2021 &\parbox[t]{4cm}{Byte bigrams\\(Top-20, norm.)} & \parbox[t]{4cm}{Malicia\\(20 fam., balanced)} & 20k & K-Means & ARI \\

\addlinespace
\cite{amorim_identifying_2021} & 2021 & \parbox[t]{4cm}{Engineered\\statics (67)} & Drive-by-download & $\sim$2k & \parbox[t]{4cm}{K-Means++\\(imwk-means)} & Euclidean \\

\addlinespace
\cite{pitolli_malfamaware_2021} & 2021 & \parbox[t]{4cm}{Static + dynamic\\(Cuckoo)} & \parbox[t]{4cm}{18 fam.\\(imbal.)} & not reported & \parbox[t]{4cm}{BIRCH\\(online)} & \parbox[t]{4cm}{CF-tree\\radius} \\

\addlinespace
\cite{yamany_new_2022} & 2022 & \parbox[t]{4cm}{Strings\\vs IAT} & \parbox[t]{4cm}{Windows\\ransomware} & not reported & \parbox[t]{4cm}{Similarity\\clustering} & Jaccard \\

\addlinespace
\cite{mosharrat_automatic_2022} & 2022 & \parbox[t]{4cm}{Android\\statistics} & \parbox[t]{4cm}{Android malware\\(RT)} & not reported & \parbox[t]{4cm}{K-Means\\(auto-$K$)} & not reported \\

% \addlinespace
% \cite{lin_anomaly-based_2022} & 2022 &
% Network flows &
% IDS traffic & not reported &
% K-Means and KNN & Euclidean \\ not malware paper - remove

\addlinespace
\cite{duraisamy_soundrapandian_mulber_2022} & 2022 & \parbox[t]{4cm}{Android\\ APK statics} & Android apps & 10.7k & \parbox[t]{4cm}{Mahalanobis\\ clustering} & Mahalanobis \\

\addlinespace
\cite{padmavathi_performance_2022} & 2022 & \parbox[t]{4cm}{Android static\\and dynamic} & \parbox[t]{4cm}{Android\\malware} & not reported & \parbox[t]{4cm}{K-Means\\DBSCAN, Aggl.} & not reported \\

\addlinespace
\cite{jureckova_classification_2024} & 2024 &
\parbox[t]{4cm}{PE statics\\and stream scores} &
\parbox[t]{4cm}{EMBER\\and unseen fam.} & not reported &
\parbox[t]{4cm}{SOM\\(online)} & \parbox[t]{4cm}{Grid\\distance} \\

\addlinespace
\cite{ilic_clustering_2024} & 2024 &
\parbox[t]{4cm}{AV-label text\\(LSA/LDA)} &
VirusTotal & 429k &
\parbox[t]{4cm}{K-Means\\(viz)} & Euclidean \\

\addlinespace
\cite{chen_android_2024} & 2024 &
\parbox[t]{4cm}{Manifest,\\API, and opcode} &
\parbox[t]{4cm}{Android\\malware} & not reported &
\parbox[t]{4cm}{K-Means, DBSCAN,\\and Spectral} & not reported \\

\addlinespace
\cite{zaborila_mace_2024} & 2024 &
\parbox[t]{4cm}{Multi-source\\artifacts} &
\parbox[t]{4cm}{Analysis\\pipeline} & not reported &
\parbox[t]{4cm}{Clustering\\engine} & not reported \\

\addlinespace
\cite{aziz_insight_2024} & 2024 &
CDR &
Mobile network & not reported &
K-Means (anomaly) & Euclidean \\

\addlinespace
\cite{mishra_discriminating_2024} & 2025 &
\parbox[t]{4cm}{Android\\temporal batches} &
KronoDroid & not reported &
\parbox[t]{4cm}{MiniBatch\\K-Means} & Euclidean \\

\bottomrule
\end{tabularx}
\footnotesize{
RT = real-time; fam. = families; imbal. = imbalanced; sel. = selected;
Aggl. = agglomerative; viz = visualization}

\caption{Summary of centroid-based clustering approaches in malware analysis.}
\label{tab:centroid-based}
\end{table*}

%%% fuzzy clustering is a centroid based

%% explain the method

%% We know that most papers have used K-means and just added some additional =feature to it.4

%% how has it evolved.

%%% main challenge selecting k value

%% fuzzy 

%% centroid based - distance metric 
%% Add metrics

%% search for availibility , keywords: 

%%2021 - Kmeans + eullcidiean , K-means + imwk, 

%% questioning the Pitoulli paper - not the best way and the results are not the best

%% no reproducibility

%% if the authors chose to create their own data they do not make it public

%% different features which makes the paper different so we talk about 

\subsection{Distribution-Based Clustering}

Unlike centroid clustering methods, distributional clustering methods analyze malware features by modeling their underlying probability distributions. Specifically, these methods, also referred to as model-based clustering, assume that the observed data are generated from a mixture of latent probability distributions with distinct parameters. When the assumed component distributions fit the empirical characteristics of the data, such methods can produce stable and informative clusters. As shown in Table~\ref{tab:distribution_based}, Gaussian mixture models are the most widely used distributional clustering approach in malware analysis. Their popularity can be attributed to their ability to estimate component membership probabilities and model correlations among features, which is particularly useful for dynamic malware analysis, where behaviors reflected in system calls, network activities, and temporal event features are often interdependent. This suitability can be further explained by the following aspects:

\paragraph{GMM as a standard clustering baseline}

Model-based clustering methods make the theoretical assumption that data is composed of a mixture of several different distributions. In malware analysis, dataset authors usually provide contextual information such as collection processes, attributes, and labels. With this basic understanding of dataset attributes, researchers can more effectively set distribution assumptions, making model-based clustering a reasonable and commonly used baseline method.

As one of the most established and widely adopted clustering models, GMM is well-known for its ability to model complex distributions and identify anomalies. Consequently, it is frequently used as a baseline method, providing a basis for comparison and reference for novel models. 
In addition, GMM is widely applied across various research domains, with active developments in its improvements and integration with other algorithms. Therefore, using GMM as a baseline for malware analysis provides a wealth of references for further improvement strategies and method selection.

Smith et al.~\cite{smith2023supervised} evaluated the performance of three representative clustering models and seven supervised learning models on malware datasets. The results showed that even in a low-correlation dataset, clustering models such as GMM were still able to produce stable results, and therefore can be used as one of the baseline clustering methods in malware analysis.

While evaluating eight baseline/state-of-the-art clustering methods on seven cross-domain datasets (covering different feature dimensions and sample sizes), Abrar et al.~\cite{abrar2023effectiveness} found that GMM, as a classic clustering model, ranked second among the eight models in terms of accuracy in the malware dataset; it also ranked second in overall performance across all seven datasets, demonstrating both accuracy and robustness.

Considering the effectiveness of GMM, Hou et al.~\cite{hou2022iot} adopt a variant of GMM, which they refer to as AE-VBGMM, to cluster reduced-dimensional IoT traffic data for anomaly detection. Their experimental results show that their model outperforms other clustering models in four evaluation metrics. 

Sometimes GMM is used as a standard model and soft clustering tool to segment complex data; each cluster is then processed with a matched detector and threshold. In anomaly traffic monitoring, this ‘Segmentation based on clustering(GMM) + threshold, then classify’ strategy can improve the performance of the resulting classifier~\cite{samunnisa2023intrusion}.

\paragraph{Efficiency of GMM on Dynamic Malware Features}

In dynamic malware analysis, consecutive activity records are often key indicators for identifying malicious behavior. For example, a series of operations, such as checking system information and then obtaining user permissions, typically indicates an attempt to elevate privileges. For such paired or consecutive features, classic clustering methods like K-means might assign independent and equal weights to each feature; however, GMM directly estimates the probability density of each parameter and explicitly models the covariance between features, thereby achieving better performance~\cite{interrante2018gaussian}. Due to its ability to characterize relationships between features, GMM demonstrates strong clustering performance in malware dynamic analysis, particularly for dynamic features such as network traffic and time-stamped logs, as we are about to describe.

% In 2018, Interrante et al.~\cite{interrante2018gaussian} conducted comprehensive evaluations of GMM clustering compared to other popular clustering models such as K-means and agglomerative clustering in malware dynamic analysis. They utilized co-occurrence matrices of system calls to represent dynamic features and conducted clustering experiments using a subset of VirusShare~\cite{virusshare2017}. The results showed that GMM significantly outperformed the other models in all evaluation metrics, and the authors also provided a theoretical explanation for why GMM is suitable for malware dynamic analysis.
% In 2019, Satoh et al.~\cite{satoh2019cause} proposed a method that uses Word2Vec~\cite{mikolov2013efficient} to establish feature representation and GMM to perform clustering based on a series of malicious DNS queries. This method classifies malicious queries based on their causes and helps administrators identify the reason for infection. In this paper, the feature of consecutive DNS queries can also be regarded as "consecutive activity records" since they reflect related activities in a short time.
In the detection of DDoS attacks, the open set recognition problem~\cite{bendale2016towards,sabeel2019evaluation} refers to the difficulty that arises when a model fails to classify new instances that are not from the distribution of its training set. Shiel et al. ~\cite{shieh2021detection} introduced a GMM model for anomaly detection, identified these new instances, and added these anomalies to the training set after being manually labeled. By applying GMM to effectively filter new instances, their method significantly improved the recall rate for the CIC-IDS2017 dataset(unknown attacks)~\cite{CICIDS2017} from 41.2\% to 99.8\%.

Log data or network traffic data that records specific times and activities can be regarded as behavior data, which is also suitable for GMM models. Paul et al.~\cite{paul2019using} collected user 
Internet logs and employed three different unsupervised models for anomaly detection, thereby avoiding malicious attacks. Even under different encoding and scaling methods, GMMs tend to have the fewest false positives.
Chapaneri et al.~\cite{chapaneri2021multi} conduct experiments on the CIC-IDS2017 network traffic dataset~\cite{sharafaldin2018toward} to detect benign and malicious traffic and further detect the type of attacks. They train multiple GMMs for each type of attack and combine them to get a comprehensive detection model, which outperforms other classification models in terms of F1 score.
In malicious attack detection, network traffic usually contains different domains, which have different sizes and do not follow a homogeneous distribution. An et al.~\cite{an2022ensemble}integrated multiple deep autoencoders to extract features from multiple dimensions and then applied a GMM model to identify anomalies. With GMM's advantages in modeling complex distributions and identifying anomalies, this method achieved significantly better performance (in terms of F1 scores) in three datasets, including KDD Cup99~\cite{kddcup99task}.

% \begin{table*}[htbp]
\begin{table}[b]
\centering
\renewcommand{\arraystretch}{1.8} %line spacing
\setlength{\tabcolsep}{4pt} 

\begin{tabularx}{\textwidth}{@{} m{0.75cm} m{1cm} m{3.8cm} m{3cm} m{1.3cm} m{1.7cm} m{3cm} @{}}
\toprule
Paper & Year & Feature & Dataset & Size & Model & Metrics \\
\midrule

% \cite{interrante2018gaussian} & 2018 & Dynamic program behavior based on co-occurrence matrices of system calls & \raggedright A subset of VirusShare\arraybackslash & 159 & GMM, Kmeans, Agglo & ARI, AMII, and FMI  \\

% \cite{satoh2019cause} & 2019 & Vectors based on co-occurrence of temporally adjacent queries & DNS query data & 372GB & GMM & F-measure  \\

% \cite{paul2019using} & 2019 & Encoded log vectors & \raggedright Web proxy log dataset \arraybackslash & 40K & GMM, Kmeans, BGMM & \parbox[t]{2.8cm}{False Positive,\\ False Negative}  \\

\cite{shieh2021detection} & 2021 & \parbox[t]{3cm}{Encoded network\\ traffic features} & \parbox[t]{2.8cm}{CIC-IDS2017,\\ FCIC-DDoS19} & 1.08M & GMM & \parbox[t]{3.3cm}{Confusion matrix,\\Ac, Pr, Re, and F1}  \\

\cite{chapaneri2021multi} & 2021 & \parbox[t]{4cm}{Log-transformed network\\ traffic features} & CIC-IDS2017 & 1.08M & GMM & Pr, Re, and F1  \\

\cite{hou2022iot} & 2022 & \parbox[t]{4cm}{3D vectors compressed\\ by autoencoders} & \parbox[t]{4cm}{IoT Network\\ Intrusion Dataset} & 2.99M & \parbox[t]{3cm}{VBGMM,\\Kmeans,\\GMM} & \parbox[t]{3cm}{Purity, Rand Index, \\F-score, and ARI}  \\

\cite{an2022ensemble} & 2022 & \parbox[t]{4cm}{Integrated features\\ from multi autoencoders} & \parbox[t]{3cm}{KDDcup99,\\ CICMalDroid20,\\ AWID2018} & \parbox[t]{1.5cm}{15K, \\ 80K,\\ 60K} & GMM & F1  \\

\cite{abrar2023effectiveness} & 2023 & \parbox[t]{4cm}{Permission and API\\ based features} & \parbox[t]{4cm}{TUNADROMD\\ and others} & 4465 & GMM, Kmeans, others & Ac  \\

\cite{smith2023supervised} & 2023 & \parbox[t]{4cm}{Vectors of malware\\ static memory features} & \parbox[t]{3cm}{CICMalMem22\\Malware-Exploratory} & \parbox[t]{2cm}{58K\\216K} & \parbox[t]{3cm}{GMM \\Kmeans \\DBSCAN} & \raggedright Consistency across clustering graphs \arraybackslash  \\

\bottomrule
\end{tabularx}
\par\vspace{2pt}
\begin{minipage}{\textwidth}
\footnotesize
\raggedright
ARI = Adjusted Rand Index, \quad
AMII = Adjusted Mutual Information Index, \quad
FMI = Fowlkes–Mallows Index, \quad
Ac = Accuracy, \quad
Pr = Precision, \quad
Re = Recall,\quad
Agglo = agglomerative clustering,\quad
BGMM = Bayesian Gaussian Mixture Models, \quad
VBGMM(AE-VBGMM) =Bayesian Gaussian Mixture Model for Autoencoder, \quad
DBSCAN = Density-based spatial clustering of applications with noise
\end{minipage}
\caption{A Brief Summary of Distribution-based Clustering}
\label{tab:distribution_based}
\end{table}

In conclusion, GMM can serve not only as a standard baseline for clustering but also as an effective approach for modeling correlations among features. By modeling feature covariances, GMM is particularly suitable for dynamic analysis tasks, including network traffic analysis, system log analysis, and behavioral sequence modeling.

% GMM AS A STAND...： the first two paragraphs are repeated, reduce a bit

% reduce the efficiency of GMM, the first paragraph needs to be reduced into 3-4 sentences, and keep the last sentence

% a conclusion paragraph at the end of this section

\subsection{Density-Based Techniques}

%% Line for continuity
Although many distribution-based clustering methods, such as GMM, are effective for modeling correlated features and dynamic behaviors, they rely on explicit assumptions about the underlying data distribution. In malware datasets where cluster shapes are irregular, densities vary, or noise is prevalent, it turns out that density-based clustering methods provide a useful alternative.

% introduce density-based clustering
Density-based clustering algorithms (DBCLAs) determine clusters by examining the spatial density of data points; objects in sparse areas are typically treated as noise or border points~\cite{kriegel2011density}. 
This principle grants DBCLAs advantages over other clustering methods, making them well-suited for malware analysis.

\paragraph{DBCLAs Handle Arbitrary Cluster Geometry}

In malware analysis, distribution patterns and clustering shapes are often influenced by technological evolution, policies, economics, and other factors that don't follow globular shapes. 
Partitional methods, such as K-means, struggle to recognize non-globular shapes, whereas DBCLAs can detect clusters of arbitrary shapes, varying densities, and scales~\cite{bhattacharjee2021survey}.
Therefore, clustering methods that don't rely on cluster shapes can better capture the non-globular and variable-density structure of malware distribution.

In density-based clustering, whether a point is in the "neighborhood" of other points determines whether that point is considered part of a cluster or considered an outlier, not part of any cluster.  This neighborhood is defined in terms of its radius, a parameter usually called $\epsilon$.
Duaa et al.~\cite{mustafa2023adaptive} introduce the Grey Wolf Optimizer (GWO) which adjusts the value of $\epsilon$ automatically. Compared to the well-known DBSCAN~\cite{ester1996density} algorithm, with manually set parameters, this approach significantly improves botnet detection accuracy on the N\_balot IoT dataset~\cite{meidan2018n}, increasing accuracy from 80\% to 98\%.

Abdullah et al.~\cite{johari2017hierarchical} are among the first researchers to explore the feasibility of HDBSCAN in malware analysis. They performed a dynamic analysis of malware in the Cuckoo sandbox, extracting both spatial and temporal features. Then, they use Jaccard distance to measure similarity between specimens and further utilize HDBSCAN for clustering. Although they didn't provide any experimental results, their study highlighted the potential value of applying HDBSCAN for malware behavior features.

%The idea of density-based clustering is also effective in malware analysis. 
Inspired by the notion of “incorporating samples below a threshold and merging clusters,” Dam et al.~\cite{dam2021unsupervised} proposed a custom, threshold-driven clustering algorithm for system call dependency graphs. They applied LSTM autoencoders~\cite{srivastava2015unsupervised} to encode system call dependency graphs and compared their clustering method with classic Ward's Method agglomerative clustering~\cite{szekely2005hierarchical}, OPTICS~\cite{ankerst1999optics}, and DBSCAN. Their approach achieves better performance in terms of metrics including Silhouette scores(SI), Adjusted Rand Index(ARI)~\cite{hubert1985comparing}, Fowlkes-Mallows Index (FMI)~\cite{fowlkes1983method}, and V-measure~\cite{rosenberg2007v}.

Recent studies have further integrated density-based clustering with hierarchical clustering, demonstrating its effectiveness for large-scale malware analysis.
As malware corpora grew, hierarchical clustering was increasingly applied at scale.  Bhandary et al. \cite{bhandary2024ransomware} applied the HDBSCAN algorithm with a custom distance metric to over 600,000 ransomware samples, clustering them based on Windows API imports and PE metadata. This was one of the first attempts to generate family-level signatures automatically at such scale. By focusing on API function imports, they produced function-level signatures resilient to polymorphic obfuscation, achieving high precision and specificity. This work demonstrated how hierarchical density-based clustering could form the basis for automated, large-scale ransomware detection pipelines.

\paragraph{Evolution-Aware Clustering}

DBSCAN uses $eps$ to define the radius of a specimen's neighborhood and $minPts$ to specify the minimum number of points threshold for a high-density region. A specimen can be regarded as a core point if it has at least $minPts$ neighbors within its $eps$-neighborhood. This algorithm uses core points as seeds, merges neighboring samples into the same cluster, and recursively expands the newly incorporated core points until no further expansion is possible. 

This seed-based expansion mechanism can be conceptually analogized to malware evolution. 
Attackers often reuse code from representative malware specimens and modify their functionality to generate numerous variants and new representative specimens~\cite{calleja2018malsource}. 
In a code-reuse feature space, such reuse can induce locally dense clusters and may form new reuse centers(representative malware specimens).

\begin{figure}%[t]
    \centering
    \includegraphics[width=0.6\textwidth]{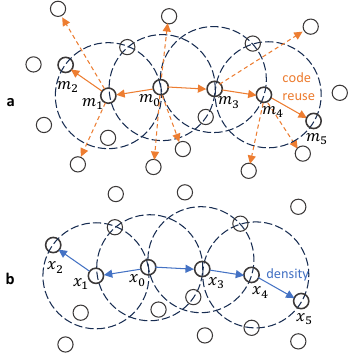}
    \caption{The cluster expansion process of DBSCAN aligns closely with the evolution pattern of malware, as malicious samples often generate numerous variants through code reuse.}
    \label{fig:dbscan}
\end{figure}

% During the development of malware, representative samples are repeatedly modified and reused, which produces multiple variants with similar structures and functionalities~\cite{calleja2018malsource}.
%\todo{this sentence is hard to understand.  I'll try to redo it.}
% When the similarity metric encodes code reuse or behavioral proximity, this evolutionary process can result in a cluster.
%, from one specimen to form a cluster according to density connectivity, aligns closely with the key feature that malware samples reuse code and generate variants.
As shown in Fig.~\ref{fig:dbscan}, DBSCAN~\cite{ester1996density} starts from $x_{0}$ and continuously expands $x_1$ to $x_5$ through density-based relationships and gradually incorporates closely related samples into the same cluster. 
In this way, DBSCAN not only captures the density structure within malware clusters but also helps identify family labels and support lineage-related analysis.
%, which makes clusters meaningful and interpretable. 

Due to these similar mechanisms between density-based clustering and malware code reuse. Liu et al.~\cite{liu2017automatic} find that shared nearest neighbor, a density-based clustering model, is helpful to detect new malware. They extract opcodes, gray-scale images, and import functions from 21740 Windows malware samples and perform malware family classification using a weighted-voting ensemble of classifiers. Then, for specimens without a clear family label, 86.7\% of new malware was identified by this density-based clustering model, according to an expert review.

DBCLA also proves highly effective in evaluating lineage analysis tasks concerning malware development and evolution~\cite{li2023automated}. In this paper, the authors remove template functions, then apply DBSCAN and OPTICS to cluster the representative functional code of each family.  Based on the selected family core function group, they further infer the lineage of malware families~\cite{li2023automated}. 

%this sentence doesn't seem to fit here, except that it uses density based clustering.  Shouldn't the liu2017automatic paper be discussed before li2023automated?
% updated

%Liu et al.~\cite{liu2017automatic} extract opcodes, gray-scale images, and import functions from 21740 Windows malware samples and perform malware family classification using a weighted-voting ensemble of classifiers. Then, for specimens without a clear family label, they apply shared nearest neighbor, a density-based clustering mode, to detect new malware. 
%According to experts' review,  86.7\% of new malware was identified by this density-based clustering model.

\paragraph{Limitation: Applicability Challenges}

Density-based clustering methods have certain limitations in their application scenarios, but with improvements, they can still effectively cluster various malware features. Padmavathi et al.~\cite{padmavathi2022performance} explore the performance of different unsupervised models in Android malware detection. They apply Principal Components Analysis (PCA)~\cite{mackiewicz1993principal} to reduce the dimensionality of 470 features in the CICMalDroid2020 dataset, then evaluate different clustering methods, including K-means, KNN, DBSCAN, OPTICS~\cite{ankerst1999optics}, hierarchical, and spectral clustering. In this paper, density-based clustering methods, DBSCAN and OPTICS, were found to fail in forming the defined clusters. To our knowledge, no existing research has systematically discussed the conditions under which DBCLAs become unsuitable for malware clustering analysis.
Similarly, Bak et al.'s research~\cite{bak2020clustering} indicates that the classic OPTICS algorithm fails to effectively cluster filtered fuzzy hash features, leaving up to 68\% of the dataset unclassified. 
The reason could be the prevalence of singletons in malware datasets. These singleton samples exhibit low similarity among other malware samples. In traditional density-based clustering algorithms, such as DBSCAN or OPTICS, singletons are often misclassified as “noise” because they fail to meet the minimum density requirements (e.g., minPts). However, in malware analysis, these singleton samples may hold significant analytical value. These singletons may, perhaps,  represent unidentified variants or even belong to entirely new malware families.%HDBSCAN
Table \ref{tab:density_based} provides a summary of all the works discussed in this section along with the features and datasets used.

% \begin{table*}[htbp]
\begin{table*}[b]
\centering
\renewcommand{\arraystretch}{1.5} %line spacing
\setlength{\tabcolsep}{4pt} 

\begin{tabularx}{\textwidth}{@{} p{0.9cm} p{0.75cm} p{3cm} p{3cm} p{1.5cm} p{3.5cm} p{2.5cm} @{}}
\toprule
Paper & Year & Feature & Dataset & Size & Model & Metrics \\
\midrule

% \cite{johari2017hierarchical} & 2017 & \parbox[t]{4cm}{Dynamic behavior\\from Cuckoo reports\\(imports, mutexes,\\ API, and so forth) } & \parbox[t]{4cm}{Virus Share~\cite{virusshare2017} \\ Das Malwerk~\cite{DasMalwerk2016} \\ Contagio Dump\footnotemark[1]\\  Malheur~\cite{Rieck_Malheur_Dataset}} & 69K & HDBSCAN & Pr, Re \\

% Can be more specific about hte feature - dynamic behavior sounds a little generic
% done

% \cite{liu2017automatic} & 2017 & \parbox[t]{4cm}{grayscale images\\Opcode n-gram\\import functions} & \parbox[t]{4cm}{Proposed\\Win PE dataset} & 21740 & SNN & Ac \\

\cite{bak2020clustering} & 2020 & TLSH hash & an IoT dataset & 12993 & \parbox[t]{4cm}{k-medoid\\OPTICS\\updated OPTICS} & Purity \\

\cite{dam2021unsupervised} & 2021 & \parbox[t]{4cm}{FastText and LSTM\\encoded SCDGs} & \parbox[t]{4cm}{Proposed\\ Win PE dataset} & 6615 & \parbox[t]{4cm}{Proposed model\\DBSCAN, OPTICS\\Ward Agglo~\cite{szekely2005hierarchical}} & \parbox[t]{4cm}{SI, ARI\\FMI\\V-measure} \\

\cite{padmavathi2022performance} & 2022 & \parbox[t]{4cm}{Static information\\behavior\\network traffic} & MalDroid2020~\cite{CICMalDroid2020} & 11598 & \parbox[t]{4cm}{KMeans, KNN\\AGNES, DIANA~\cite{kaufman2009finding}\\ DBSCAN, OPTICS\\ Spectral~\cite{von2007tutorial}} & \parbox[t]{4cm}{RI, MIBS\\FMS, SC\\ CH} \\

\cite{mustafa2023adaptive} & 2023 & IoT network traffic & N-BaIoT~\cite{meidan2018n} & 7M & \parbox[t]{3cm}{DBSCAN-GWO} & Ac, Pr, Re, DR \\

\cite{li2023automated} & 2023 & Enhanced ACFG & BODMAS~\cite{yang2021bodmas} & 57K & \parbox[t]{4cm}{DBSCAN, KNN\\OPTICS} & \parbox[t]{4cm}{Ac, Coverage\\Macro Pr\\ Macro Re} \\

\bottomrule
\end{tabularx}

\par\vspace{2pt}
\begin{minipage}{\textwidth}
\footnotesize
\raggedright

Ac = Accuracy, \quad
Pr = Precision, \quad
Re = Recall,\quad
SCDG = System Call Dependency Graphs, \quad
SI = Silhouette score,\quad
ARI = Adjusted Rand Index,\quad
FMI = Fowlkes-Mallows Index,\quad
ACFG = Attribute Control Flow Graph,\quad
SNN = Shared Nearest Neighbor,\quad
MIBS = Mutual Information Based Score,\quad
RI = Rand Index,\quad
FMS = Fowlkes-Mallows Score,\quad
SC = Silhouette Coefficient,\quad
CH = Calinski-Harabasz Index
\end{minipage}
\caption{A Brief Summary of Density-based Clustering}
\label{tab:density_based}
\end{table*}

% \footnotetext[1]{\url{https://contagiodump.blogspot.com/}}

% [15] should be included in density-based(copy pasted, but we need to refrize to make it smooth)
% updated

\subsection{Hierarchical Clustering}

%%A combination of papers from the past two years! Yippee! 

%% Add of the hierarchical adds on to density

% \begin{figure}[t]
%     \centering
%     \includegraphics[width=\columnwidth]{clusteringfigures/Hierarchical-based-cluster.png}
%     \caption{(Placeholder) Representation of Hierarchical-based}
%     \label{fig:hierarchical-based}
% \end{figure}

In contrast to density-based methods that emphasize local structure and noise robustness, hierarchical clustering approaches focus on organizing malware samples into multi-level relationships that capture lineage, similarity, and evolution.

Hierarchical clustering methods have become increasingly prominent in malware analysis due to their ability to organize samples at multiple levels of granularity. Unlike flat clustering approaches, hierarchical methods produce nested structures that capture relationships between malware samples, families, and variants, enabling both exploratory analysis and operational use. This makes them particularly well suited for understanding malware evolution, generating interpretable signatures, and supporting detection in dynamic and large-scale environments.

In recent years, hierarchical clustering in this area used dendrograms to visualize family structures and evolutionary relationships. For example, \cite{naveen_kumar_robust_2024} combined stacking classifiers with dendrogram visualization to build a robust malware classification system. Their approach grouped malware samples according to ancestral relationships, helping analysts track variant evolution while enhancing classification accuracy. This line of work highlighted how hierarchical clustering could complement classification by offering both precision and interpretability.

%shifting this to density based

% As malware corpora grew, hierarchical clustering was increasingly applied at scale.  Bhandary et al. \cite{bhandary2024ransomware} applied the HDBSCAN algorithm with a custom distance metric to over 600,000 ransomware samples, clustering them based on Windows API imports and PE metadata. This was one of the first attempts to generate family-level signatures automatically at such scale. By focusing on API function imports, they produced function-level signatures resilient to polymorphic obfuscation, achieving high precision and specificity. This work demonstrated how hierarchical density-based clustering could form the basis for automated, large-scale ransomware detection pipelines.

Scalability concerns have also driven research into efficient similarity approximations for hierarchical clustering. S. et al.~\cite{s_hierarchical_2025} evaluated hierarchical clustering over opcode n-gram representations derived from the SOREL-20M corpus, comparing Jaccard, MinHash, and SuperMinHash distances. Their results demonstrated that hierarchical methods can remain effective on large malware datasets when paired with appropriate similarity approximations, balancing clustering fidelity and computational cost.

Beyond static analysis, hierarchical concepts have been extended into behavioral and profiling frameworks. Ovayan et al.~\cite{ovayan_novel_2024} proposed Hierarchical Threat Signal Profiling (HTSP), which models ransomware behavior across multiple runtime layers, including system activity, encryption behavior, and network anomalies. Rather than clustering binaries directly, HTSP uses hierarchical structure to support adaptive, real-time detection. This reflected a shift toward autonomous profiling, where clustering provides the structural backbone for adaptive detection.

Opcode analysis has similarly benefited from hierarchical organization. Gu et al.\cite{gu_hierarchical_2024} developed a hierarchical K-nearest neighbors approach for ransomware detection, combining hierarchical structure with instance-based classification on opcode sequences. Their model improved detection accuracy and robustness, particularly for noisy or high-dimensional opcode data. Ko et al. (2025)\cite{ko_n-gram_2025} further demonstrated that hierarchical clustering can serve as an effective preprocessing step by structuring opcode n-gram features before applying convolutional neural networks. These approaches show how hierarchical clustering is increasingly used not only as an end goal for family grouping but also as a bridge into feature engineering for advanced ML pipelines.

Recent work has also explored behavioral entropy–based frameworks that push the hierarchical idea even further. Aqazi et al.\cite{aqazi_hierarchical_2024} proposed Hierarchical Behavioral Entropy Mapping (HBEM), which aggregated entropy deviations across files, memory, and network behaviors into a hierarchical decision framework. Similarly, Kurov et al.\cite{kurov_deciphering_2024} used hierarchical anomaly-based encryption analysis, designed to distinguish malicious from legitimate encryption processes with high precision. Both approaches highlight how hierarchical decomposition of system behaviors enables detection that is resilient to obfuscation, polymorphism, and mimicry, while maintaining scalability for real-time environments.

Taken together, these studies, as summarized in Table \ref{tab:hierarchical-based}, illustrate a thread of interest in hierarchical clustering in malware research. Initially used for visualization and family grouping, hierarchical methods have progressed toward large-scale static clustering and are now embedded within dynamic and hybrid profiling frameworks. This progression reflects a broader shift from purely exploratory analysis toward adaptive, interpretable, and operationally scalable malware detection systems.

% Looking forward, several directions appear promising. Hierarchical clustering could be further integrated with deep learning models such as CNNs and transformers, where cluster structures guide representation learning while retaining interpretability. Temporal hierarchical models that capture stage-wise ransomware execution patterns may offer even finer detection granularity. Entropy and anomaly-based clustering could be fused with system log or network telemetry analysis to create resilient multi-layer defenses. Finally, dendrograms and hierarchical maps remain powerful interpretability tools, suggesting that explainable clustering will play an important role in analyst-centric security systems.

In practical terms, hierarchical methods enrich malware defense in two ways. First, they improve explainability: dendrograms, entropy maps, and hierarchical profiles give cybersecurity professionals interpretable insights into how malware samples are related, making detection decisions more transparent. Second, they support adaptability: as malware families fragment, rebrand, or evolve, hierarchical models can incorporate new behaviors while preserving lineage information, aiding both research and operational response. This positions hierarchical clustering as not just a classification tool but a mechanism for tracking evolution, guiding signature generation, and enabling hybrid integration with deep learning or distributed frameworks.

%% leading paragraph to the next part - mixed type - can be added in the next subsection as well

Although hierarchical clustering has proven effective for organizing malware into interpretable family structures and modeling evolutionary relationships, many modern malware datasets do not conform to a single homogeneous feature space. In practice, malware samples are often described by a mixture of static, dynamic, behavioral, textual, and network-derived features, each with different statistical properties and scales. This heterogeneity limits the applicability of clustering methods that assume uniform feature representations. As a result, recent research has increasingly explored mixed-type clustering approaches, which are explicitly designed to handle heterogeneous feature spaces and integrate multiple views of malware behavior within a unified clustering framework

% \begin{table*}[htbp]
\begin{table*}[b]

\centering
\renewcommand{\arraystretch}{1} %line spacing
\setlength{\tabcolsep}{4pt} 

\begin{tabularx}{\textwidth}{@{}p{0.7cm} p{0.7cm} p{3.75cm} p{3cm} p{1cm} p{2.5cm} p{3cm} @{}}
\toprule
Paper & Year & Feature & Dataset & Size & Model & Metrics \\
\midrule

\cite{naveen_kumar_robust_2024} & 2024 & 56 features Static  & \raggedright CIC-MalMem-2022 \arraybackslash & 58,596 X 56 & \parbox[t]{4cm}{LR+SVM\\ +KNN+NB} & Ac, FPR, and DR  \\

\cite{bhandary2024ransomware} & 2024 & \parbox[t]{4cm}{DLL imports\\ and PE features} & \raggedright A subset of SOREL-20M \arraybackslash & 1.1M & HDBSCAN & \parbox[t]{4cm}{PPV, TPR, TNR,\\ FPR, FNR, NPV}  \\

\cite{s_hierarchical_2025} & 2025 & Dissemble Opcodes & \raggedright A subset of SOREL-20M \arraybackslash & 159 & \parbox[t]{4cm}{GMM, Kmeans\\ Agglo} & ARI, AMII, and FMI  \\

\cite{ovayan_novel_2024} & 2024 & \parbox[t]{4cm}{runtime signals, network\\ traffic, encryption behaviors} & \raggedright private \arraybackslash & n/r & HTSP & Ac, precision, recall, F1-score  \\

\cite{gu_hierarchical_2024} & 2024 & Opcodes & \raggedright -(random) \arraybackslash & 284 & hierarchical-KNN  &  Ac, precision, recall \\

\cite{ko_n-gram_2025} & 2025 & Opcodes & \raggedright KISA \arraybackslash & 159 & CNN & Ac, precision, recall, F1-score  \\

\cite{aqazi_hierarchical_2024} & 2024 & entropy & \raggedright \arraybackslash -(random) & n/r & HBEM & acc, precision, recall, FPR  \\

\cite{kurov_deciphering_2024} & 2024 & encryption & \raggedright -(random) \arraybackslash & 1 M & GMM, Kmeans, Agglo & \parbox[t]{4cm}{Ac, FPR, FNR,\\ processing speed}  \\

\bottomrule
\end{tabularx}
\par\vspace{2pt}
\begin{minipage}{\textwidth}
\footnotesize
\raggedright
Ac = Accuracy, \quad
Pr = Precision, \quad
Re = Recall,\quad
FPR = False Positive Rate, \quad
FNR = False Negative Rate, \quad
HDBSCAN = Hierarchical Density-based spatial clustering of applications with noise
n/r = not reported
\end{minipage}
\caption{A Brief Summary of Hierarchical-based Clustering}
\label{tab:hierarchical-based} 
\end{table*}

\subsection{Mixed clustering methods}
% citations need to be fixed
% This section discusses the applications of clustering methods that are impleted in combination with other clustering or machine learning methods.
Although different clustering algorithms have helped malware analysts answer different questions about a given collection of malware, there has also been a growing interest in the community for implementations of combinations of clustering algorithms, sometimes with other clustering algorithms and in some situations with other machine learning algorithms. Clustering Ensembles, as these approaches are called, usually have two or more algorithms that work in tandem to address certain challenges that no single algorithm can handle. 
%\todo{does the following paragraph still make sense?}
For example, Maazalahi et al. \cite{maazalahi_hybrid_2024} propose the use of Genetic Algorithms (GA) along with K-means clustering to detect malware.  To be specific, their method uses GA to find an initial set of cluster centers for the K-means algorithm to start with. This method of using another algorithm before clustering was found to improve the accuracy of malware identification and detection as shown by Taha et al. \cite{taha_hybrid_2021} where they demonstrate the use fuzzy clustering along with LightGBM.

Unsupervised machine learning algorithms commonly used for clustering ensembles include hierarchical clustering, K-means, and K-mediods. There is a growing interest in methods that are combinations of hierarchical, K-means, and K-mediods, either with each other or with other machine learning methods \cite{ye2010automatic}, \cite{han_parameter-free_2009}, \cite{ali_scalable_2020} \cite{gu_hierarchical_2024}. Dener's work \cite{dener_clustering-aided_2023} proposes a model that predicts malware by using K-means clusters and classifiers. The authors classify each cluster with different classification models for better prediction, with the results from each classification method compared with other results based on their accuracy and F1 scores. This combination is able to provide better predictions than clustering or classifications separately.

Bhattarai et al. \cite{bhattarai_heal_2024} shows an implementation with hierarchical and fuzzy clustering with a malware dataset, demonstrating how this combination can be used to improve on document classification and reducing hallucinations in LLM outputs.

We note that many if not most the works discussed in this section used datasets made up of Android malware. This could be due to clustering ensembles performing better with Android malware and questions could be asked about why other malware types don't seem to perform as well with these methods. A brief overview of this has been provided in Table \ref{tab:mixed}.

There are also methods where authors have used clustering algorithms with deep learning techniques. This is discussed in the following section.

%Most applications of these methods are for android malware
%• Some methods use a combination of clustering and deep learning or a supervised learning

% \begin{table*}[!htbp]
\begin{table*}[b]
\centering
\renewcommand{\arraystretch}{1.5} %line spacing
\setlength{\tabcolsep}{4pt} 

\begin{tabularx}{\textwidth}{@{}p{0.75cm} p{0.7cm} p{3cm} p{2.8cm} p{2.3cm} p{2.5cm} p{3.5cm} @{}}
\toprule
Paper & Year & Feature & Dataset & Size & Model & Metrics \\
\midrule
\cite{han_parameter-free_2009} & 2009 & Not specified & \parbox[t]{4cm}{Kingsoft\\AVLab} & \parbox[t]{4cm}{3198 samples\\(3 datasets)} & \parbox[t]{4cm}{Hierarchical\\K-means} & F1 score \\

\cite{ye2010automatic} & 2010 & \parbox[t]{4cm}{Function based\\ instruction} & \parbox[t]{4cm}{Kingsoft\\AVLab} & No Infor. & \parbox[t]{4cm}{Hierarchical\\K-mediods} & F1 score\\

\cite{ali_scalable_2020} & 2020 & Not specified & VirusTotal & No Infor. & \parbox[t]{4cm}{K-Means\\Hierarchical} & Purity score  \\

\cite{xiao_hybrid_2021} & 2021 & Static features & CICAndMal2017 & 7067 & New model & AUC, Pr, Re, and F1\\

\cite{taha_hybrid_2021} & 2021 & App permissions & Drebin & 129,000 & \parbox[t]{4cm}{Fuzzy clustering\\LightGBM} & Pr, Re \\

\cite{dener_clustering-aided_2023} & 2023 & Not specified & \parbox[t]{4cm}{Kaggle\\EMBER2018\\BODMAS} & \parbox[t]{4cm}{216k\\134k\\1M} & \parbox[t]{4cm}{K-Means\\multi classifiers} & \parbox[t]{4cm}{Ac\\F1 score}\\

\cite{maazalahi_hybrid_2024} & 2024 & Not specified & CICMaldroid2020, Andro-Autopsy & No Infor. & \parbox[t]{4cm}{Genetic algo.\\K-means} & \parbox[t]{4cm}{Pr, Re,\\and F-measure}\\

\cite{gu_hierarchical_2024} & 2024 & Not specified & No Infor. & No Infor. & Hierarchical, KNN & \\

\bottomrule
\end{tabularx}
\par\vspace{2pt}
\begin{minipage}{\textwidth}
\footnotesize
\raggedright
Ac = Accuracy, \quad
Pr = Precision, \quad
Re = Recall,\quad
\end{minipage}
\caption{A Brief Summary of Mixed Clustering}
\label{tab:mixed}
\end{table*}

\subsection{AI and Deep Learning based Clustering}
In recent years, AI- and DL-based methods, especially LLMs, have been widely used to optimize and assist clustering. However, malware analysis research that combines LLMs and clustering methods remains limited.
Wolsey's survey \cite{wolsey_state---art_2022} provides a comprehensive analysis of AI-based malware detection techniques. Those authors also highlight the use of AI to create malware, which reduces the time a malware author spends creating a new malware specimen and, similarly, how AI-based methods can be used to detect and identify malware.

Rezaeu et al. \cite{rezaei_pe_2021} show how clustering can be incorporated with deep learning techniques to get better results for the detection of malware. Ng's work \cite{ng_static_2019} proposes a deep learning ensemble for malware detection that uses clustering on a static dataset. At the time of its publication, this was among the earliest works to show that this approach was both practical and effective.

It has been observed that machine learning methods, such as Deep Neural Networks, show potential for improving malware detection when combined with clustering methods. Rathore et al. \cite{rathore_detection_2021} present a comparison of the performance of classical machine learning methods and DNNs integrated with clustering for detecting Android malware. While their findings show that DNN and Autoencoder don't perform as well as expected, there is potential to improve these methods with additional resources as they become available. Similar work is presented by Patil et al~\cite{patil_malware_2020}.

\subsubsection{The use of LLMs in clustering}
Recent advancements in Large Language Models (LLMs) have enabled their use in clustering and, more broadly, in malware generation and analysis. Al Karaki's survey \cite{al-karaki_exploring_2024} provides an in-depth study of the rising popularity of LLMs for malware detection over the last few years and describes the growing interest in using LLMs to analyze malware.  For example, the MACE framework~\cite{zaborila_mace_2024}  tests control flow graphs and strings and clusters them using DBSCAN, GMM and Hierarchical clustering. The features are embedded using BERT and two other models. Results show that the framework can achieve over 97\% accuracy with most clustering algorithms. BERTDetect \cite{ranaweera_bertdetect_2025}, an approach that uses topic modeling based on Latent Dirichlet Allocation (LDA) and K-means, as well as ChatGPT embedded with K-means, to detect Android malware based on their text description and API calls. They present evidence that BERTopic performs better than LLM-based GPT embeddings.
%\todo{is this previous sentence still correct?}
% Yes
To our knowledge, the use of LLMs for malware clustering is an emerging field. There are numerous questions that could be answered and addressed with LLMs. One such problem is identifying malware families based on their assembly instructions. This knowledge could also be used to establish better guardrails for LLMs to prevent the generation of malicious code.

Oak's \cite{oak2019malware} work models the dynamic behavior data of Android applications as low-level natural language and employs BERT and LSTM for malware classification. In an extremely imbalanced dataset, the BERT model achieves outstanding performance. This work introduces a method that uses language models to describe malware features and proves the effectiveness of natural language models for malware analysis.

MaLAware applies five large language models to convert Cuckoo Sandbox reports into human-readable summaries of malicious behavior, while evaluating the consistency between these generated summaries and manually authored reference descriptions (ground truth)~\cite{saha2025malaware}. Although the authors didn't propose a large model-based clustering method, clustering can still be combined with LLMs, for example, to further classify and categorize behavior summaries generated by large models.

% =========================================================
% Dataset, Discussion, Conclusion
% =========================================================

\section{Malware Datasets and Features}

This section reviews publicly available malware datasets commonly used in clustering-based malware analysis and discusses the types of features derived from these datasets. Our survey of clustering-related literature indicates that a number of datasets are available for public research use. These datasets typically consist of curated and documented collections of benign and malicious binaries, often accompanied by metadata such as PE headers, family labels, system calls, and other static or dynamic attributes.

Table~\ref{tab:dataset_list} lists representative malware corpora that have been published or actively used in research since 2021. Despite the availability of public datasets, our review shows that many studies continue to rely on custom-collected malware samples tailored to specific experiments, or on datasets maintained by private organizations. Although this practice allows authors to carefully select samples aligned with their research objectives, it also limits reproducibility and cross-study comparison.

An additional trend observed in recent years is the growing availability of Android malware datasets, which now appear more frequently in clustering studies than in earlier work. These datasets often provide richer contextual information, including application metadata, temporal attributes, and family annotations, enabling broader exploration of clustering methods beyond Windows-based malware.

Although specially collected datasets offer flexibility, public datasets provide researchers with diverse and large-scale collections of malware that include standardized features and labels. Such datasets play a critical role in enabling reproducible evaluation and comparative analysis, particularly for clustering methods that are sensitive to data distribution and feature representation.

%\begin{itemize}
%    \item Listing some recent malware datasets. their size and OS.
%\end{itemize}

\begin{table*}[b]
\centering
\footnotesize
\setlength{\tabcolsep}{4pt}
\begin{tabularx}{\textwidth}{p{0.20\textwidth} p{0.08\textwidth} p{0.10\textwidth} X p{0.12\textwidth}}

% \onecolumn
% \begin{tabular}{lllll}
% \caption{Papers read}
\toprule
\textbf{Dataset Name} & \textbf{Year} & \textbf{OS} & \textbf{Features} & \textbf{Dataset Size} \\
\midrule

SOREL-20M\cite{harang2021sorel} & 2021 & Windows & PE & 20M\\
Ransmap\cite{hirano_ransmap_2025} & 2025 & Windows & Storage and Memory access patterns & NI \\
AndroOBFS\cite{kumar_androobfs_2022} & 2022 & Android & Time tags, family information & 16279 \\
BODMAS\cite{yang_bodmas_2021} & 2021 & Windows & PE & NI \\
EMBER2024\cite{joyce_ember2024_2025} & 2025 & Multi-sys & Previously undetected samples & 3.2M \\
MOTIF\cite{joyce_motif_2023} & 2023 & Windows & Malware tagged with families & 3095 \\
MalTracker\cite{yu_maltracker_2024} & 2024 & Multi-sys & Malicious packages & 229 \\
RMVDroid\cite{wang_rmvdroid_2019} & 2019 & Android & App metadata and families & 9000 \\
Kronodroid\cite{guerra-manzanares_kronodroid_2021} & 2021 & Android & Static, dynamic features and families & 41000 \\

% don't want to cite this one
% RIAD \cite{11012104} & 2025 & Windows & Windows Event Logs & 20 \\

RIAD-ES \cite{bhandary2025ransomwareAD} & 2025 & Windows & Windows Event Logs & 103\\ 

MASCOT\cite{li2025mascotanalyzingmalwareevolution} & 2025 & Winsows & Source files & 6032 \\

% the following datasets are from distribution-based section (only representative datasets)
CIC-IDS2017~\cite{sharafaldin2018toward} & 2017 & Multi-sys &Network traffic data & NI \\ 

CIC-DDoS2019~\cite{CICDDoS2019} & 2019 & Multi-sys & Network flow characteristics(DDoS) & 12.8M \\

IoT network intrusion~\cite{kang2019iot} & 2019 & IoT &IoT abnormal behavior Traffic characteristics & 2.99M \\

% KDDCup1999~\cite{KDDCup1999} & 1999 & Multi-sys &41 connection-based network traffic features & 18M \\

\bottomrule
\end{tabularx}
%\begin{minipage}{\textwidth}
%\footnotesize
%\raggedright
%NI = No Information, \quad
%S = Samples \quad
%\end{minipage}
\caption{List of recent malware datasets available at present.}\label{tab:dataset_list}
\end{table*}
% \end{tabular}
%% switch back to two column

%%%short forms NI - no information
%%%S -samples

% \twocolumn

%% I propose to write dataset collection as 
%% (a) Labeled Static Feature Datasets
%% (b) Malware Repositories and Meta-Corpora
%% (c) Dynamic and Behavioral Datasets

% Do you mean we rewrite of the only the table? 

%%% Features part

\begin{table*}[b]
\centering

\scriptsize
\setlength{\tabcolsep}{4pt}
\renewcommand{\arraystretch}{1.15}

\begin{tabular}{lccccccccccccccc}
\toprule
\textbf{Feature} &
\textbf{Centroid} &
\textbf{Hierarchical} &
\textbf{Distribution} &
\textbf{Density}
 \\
\midrule

Bytes              & 1 & - & - & - \\
Strings            & 1 & - & - & - \\
IAT                & 1 & - & - & - \\
API Calls          & - & 1 & 1 & - \\
AV Labels          & - & - & - & - \\
Registry Data      & 1 & - & - & - \\
Opcodes            & - & 3 & - & - \\
Network Data       & - & 1 & 4 & 1 \\
PE Features        & - & 1 & - & - \\
CDR                & 1 & - & - & - \\
System Calls       & - & - & 1 & - \\
Memory             & - & - & 1 & - \\
DLL Imports        & 1 & 2 & - & - \\
Runtime Encryption & - & 1 & - & - \\
Entropy            & - & 1 & - & - \\
TLSH               & - & 1 & 1 & 1 \\
\bottomrule
\end{tabular}
\caption{Feature usage across clustering methods in malware analysis}
\label{tab:features-by-method}
\end{table*}

Table ~\ref{tab:features-by-method} provides a consolidated view of how different feature types have been utilized across clustering methods in malware analysis. Rather than focusing on individual studies, the table aggregates feature usage patterns at the method level, highlighting which feature representations are most commonly paired with specific clustering paradigms. This perspective helps clarify how methodological choices are often intertwined with feature availability, data modality, and analytical goals.

Several trends emerge from this comparison. Static features such as bytes, strings, import tables, and PE metadata are most frequently associated with centroid-based approaches, reflecting their compatibility with fixed-length vector representations. In contrast, opcode-based features appear predominantly in hierarchical methods, where structural relationships and sequential patterns can be better preserved. Dynamic and behavioral features, including network data, runtime encryption signals, and entropy measures, are more often coupled with probabilistic or specialized hierarchical frameworks, which explicitly model feature correlations or layered behaviors.

By presenting feature usage in this aggregated manner, the table emphasizes that clustering performance is not driven by algorithm choice alone, but by the alignment between feature characteristics and clustering assumptions. It also reveals underexplored combinations, such as richer behavioral features paired with density-based or mixed-type clustering, suggesting opportunities for future work. Overall, this summary complements the method-centric organization of the survey by making explicit how feature design and clustering strategy co-evolve in malware analysis research.

\section{Discussion}

This paper surveys six major approaches to clustering  applied to malware analysis: centroid-based, density-based, distribution-based, hierarchical, hybrid-based, and AI-driven techniques. While prior work largely reports usage trends, our analysis highlights how these methods differ with respect to concept drift, outlier handling, dataset scale, and feature representation, which are key factors for practitioners selecting an approach.
Centroid-based algorithms such as K-means remain the most widely used due to their scalability and compatibility with vectorized malware features. These methods perform well on large-scale datasets with relatively clean structure and well-separated clusters. However, they are sensitive to outliers, require the number of clusters to be specified in advance, and do not naturally adapt to concept drift without extensions such as incremental updates or ensemble integration.
Density-based and hierarchical methods have seen increased adoption since around 2020, reflecting a shift toward evolution-aware clustering. Density-based methods are particularly effective in datasets with high noise, irregular cluster shapes, and varying densities, making them suitable for identifying outliers and novel malware variants. However, they can struggle when data is sparse or when density parameters are not well aligned with the dataset, sometimes resulting in a large number of unclustered samples.
Hierarchical clustering provides a complementary perspective by exposing structure at multiple levels of granularity. This makes it well suited for analyzing concept drift and lineage relationships, where clusters evolve over time. It also offers strong interpretability, allowing analysts to trace how malware families split and merge. The primary limitation is scalability, as computational cost increases significantly with dataset size, making it less practical for very large corpora without approximation or hybridization.
Distribution-based methods, though fewer in number, occupy an important niche in scenarios where feature correlations and uncertainty must be modeled explicitly. Approaches such as Gaussian Mixture Models are particularly effective for dynamic and temporally structured data, including network traffic and event logs. These methods can capture richer feature relationships than centroid-based approaches, but they are typically more computationally intensive and may require careful tuning in high-dimensional settings.
Hybrid-based approaches combine multiple clustering paradigms to balance trade-offs across scale, robustness, and adaptability. These methods are especially useful when datasets exhibit mixed characteristics, such as large scale combined with noise or evolving behavior. Our survey shows that hierarchical components are frequently included in ensembles to improve structural understanding, while centroid-based methods are used to improve efficiency.
More recently, AI-driven and LLM-based clustering approaches have emerged, leveraging learned representations to handle high-dimensional and semantically rich features. These methods show promise in improving clustering quality and interpretability, particularly in complex malware datasets. However, their behavior under concept drift and adversarial conditions remains under-explored, and current evidence suggests they are best used in combination with traditional clustering techniques.
Overall, the surveyed literature suggests that no single clustering method is optimal across all scenarios. Instead, the choice of method should be guided by dataset characteristics: centroid-based methods for large and well-structured data, density-based methods for noisy and evolving datasets, hierarchical methods for interpretability and drift analysis, and hybrid or AI-driven approaches when combining multiple strengths is required.

% Will find a better place to add it

% While numerous works discussed in this survey explore novel methods to improve accuracy of clustering methods, we observe very little work done in addressing common challenges encountered during clustering. This includes problems such as handling outliers, drift and scalability. There is very limited work in malware analysis methods that use clustering that focuses on detecting and handling outliers in a meaningful way, this can lead to some malware being not labeled right or treated as noise when they could represent a new family or class of malware, different binary structure or simply labeled incorrectly.  ~\cite{swargam_analyzing_2025} tracks how often various clustering algorithms label malware samples as noise, showing the value of outliers and the need for further study in this topic.

% With the recent rise in LLMs and their capabilities, threat actors using them to develop and modify malware has aided them in speeding up the development cycle of malware. 

%% rewritting the above well-written paragraph because I am blind and I did not see it

Although malware clustering has matured significantly over the past two decades, several fundamental research challenges remain unresolved.
\paragraph{Outlier Analysis and Novel Malware Discovery} Outlier handling remains one of the least studied aspects of malware clustering. While most clustering studies focus on cluster quality and malware family separation, comparatively few investigate the samples classified as noise. Recent work has demonstrated that clustering outliers may correspond to mislabeled samples, structurally unique binaries, or potentially novel malware variants, suggesting that outlier analysis can provide valuable insights beyond traditional clustering evaluation metrics. However, the broader role of outlier in malware triage, variant discovery, and threat intelligence remains largely unexplored\cite{swargam_analyzing_2025}. \cite{swargam_analyzing_2025}

\paragraph{Concept Drift and Malware Evolution} Malware continuously evolve through code reuse, obfuscation, and behavioral adaptation. While hierarchical and density-based approaches have demonstrated potential for tracking evolutionary relationships, there remains no consensus on how clustering systems should adapt to long-term behavioral drift. Open questions include determining when a variant should be considered a new cluster, how cluster structures evolve over time, and how clustering models can remain effective without frequent retraining. 

\paragraph{Scalability to Operational Datasets} Most malware clustering studies evaluate methods on datasets containing thousands or tens of thousands of samples. In contrast, operational environments may encounter millions of files and telemetry records daily. This raises important questions regarding the scalability of distance computations, memory requirements, and cluster maintenance. Future work should explore distributed, streaming, and approximate clustering techniques capable of operating at operational scale while maintaining clustering quality.

\paragraph{Interpretability and Representative Specimens} Clustering algorithms typically group malware samples without providing explanations for why those samples belong together. As a result, analysts often face challenges interpreting cluster meaning and identifying representative samples. Future research should investigate methods for automatically characterizing cluster behavior, identifying representative specimens, and generating analyst-friendly descriptions that support threat intelligence workflows.

\paragraph{AI-Driven Malware and AI-Driven Clustering} Recent advances in large language models and generative AI have introduced new opportunities for malware analysis but also new challenges. While AI-driven clustering approaches have shown promise in handling complex feature spaces, their computational requirements remain substantial. At the same time, the emergence of AI-generated malware raises important questions regarding attribution, clustering robustness, and behavioral similarity. Can clustering methods identify patterns unique to AI-generated malware? Can lightweight AI models provide practical alternatives to large-scale architectures? These questions remain largely unexplored.

\paragraph{Adversarial and Anti-Clustering Malware} Traditional malware evasion techniques target signatures, sandboxes, and machine learning classifiers. An emerging concern is whether malware can be intentionally designed to evade clustering itself. Obfuscation, polymorphism, feature manipulation, and cluster-poisoning strategies may reduce clustering effectiveness or fragment malware families into misleading groups. Understanding how clustering algorithms perform under adversarial conditions represents an important direction for future research.

\paragraph{Dataset Diversity and Benchmarking}The majority of publicly available malware clustering datasets focus on Windows malware, while comparatively fewer resources exist for Android, IoT, embedded systems, Linux, and cross-platform malware. This limits our understanding of how clustering techniques generalize across different environments. Future work should explore whether platform-specific representations are necessary and establish standardized benchmark datasets that facilitate reproducible evaluation across malware ecosystems.

\paragraph{Emerging Clustering Paradigms}Finally, most malware clustering research continues to rely on established clustering frameworks such as centroid-based, density-based, and hierarchical approaches. Emerging paradigms, including graph-based, self-supervised, foundation-model-based, and quantum-inspired clustering techniques, remain largely unexplored within malware analysis. Determining whether these approaches provide meaningful advantages over existing methods presents an important opportunity for future investigation.

%conclusion in some ways?
%Therefore, despite substantial advances in malware clustering, several challenges reamin unresolved. Future research must address scalability to operational-scale datasets, robust handling of outliers and evolving malware behaviors, and improved interpretability of clustering techniques, and increasingly diverse computing platforms introduce additional complexity that existing clustering frameworks were not designed to address. These challenges suggest that future malware clustering research will extend beyond improving clustering accuracy alone and toward developing adaptive, explainable, and resilient systems capable of operating in rapidly evolving threat environments.

\section{Conclusion and Future Work}

This survey reviewed recent research on malware clustering and organized the literature into seven methodological categories: centroid-based, distribution-based, density-based, hierarchical, mixed-type, fuzzy, and AI-driven approaches. Across these categories, we summarized how clustering has been used to group unknown samples, infer family structure, and support downstream tasks such as signature generation, anomaly discovery, and behavioral profiling. By consolidating prior work into method-specific summaries and comparative tables, this survey clarifies what has been explored, what assumptions are commonly made, and where empirical gaps remain.
A recurring theme across the literature is that clustering outcomes are often driven as much by feature design and data availability as by the clustering algorithm itself. Static representations such as bytes, strings, PE metadata, and imports remain common due to ease of extraction and scalability, while behavioral features such as system calls, network telemetry, and runtime signals are increasingly used when interpretability and evolution-aware grouping are important. The feature summary complements the method-centric organization by making explicit which feature types are most frequently paired with particular clustering paradigms, and it also exposes underexplored pairings that may be valuable in future evaluations.
This survey also highlights persistent challenges in reproducibility and cross-paper comparison. Although several public datasets and corpora are available and growing, many studies still rely on custom or private collections, and experimental details such as preprocessing, labeling assumptions, temporal scope, and evaluation metrics are not consistently reported. Since clustering is sensitive to feature scaling, class imbalance, label noise, and temporal drift, these inconsistencies limit the community’s ability to compare methods fairly and to assess generalization across malware ecosystems.

% %% Future Work

Future work can build on these findings in several directions. First, the community would benefit from more standardized evaluation practices, including clearer reporting of preprocessing pipelines, distance metrics, hyperparameters, and both internal and external clustering validation measures. Second, broader benchmarking across feature modalities is needed, especially for mixed-type and behavioral representations that more closely reflect operational telemetry. Third, temporal evaluation should become more common, since malware evolution and drift can change the geometry and density structure that clustering methods rely on. Finally, AI-driven approaches, including representation learning and LLM-based embeddings, are likely to shape future malware clustering research by enabling richer feature extraction and more scalable analysis. However, these methods also introduce new questions about interpretability, stability across model updates, and reproducibility, which motivates careful evaluation and open baselines.
Overall, this survey provides a consolidated view of malware clustering methods, datasets, and feature usage patterns in recent work. By identifying methodological and empirical gaps, it aims to support more rigorous, reproducible, and operationally relevant clustering research as malware continues to evolve in scale and complexity.

\section{Author Contributions}
R.S, P.B and B.L were responsible for the collection, investigation of the research works, as well as drafting the manuscript. C.N was responsible for reviewing and editing the manuscript. All authors have read and share full responsibility for the decision to submit this work for publication, and agree to be accountable to all aspects of this work, ensuring any questions related to any part of this work are appropriately investigated and resolved. 
\section{Declaration of Competing Interests}
The authors declare no competing interests.
\section{Funding Declaration}
Funding: Not Applicable

Correspondence and requests for materials should be addressed to Raguvir S.

% =========================================================
% Bibliography
% =========================================================

\bibliographystyle{cas-model2-names}

\bibliography{clustering,clusteringlocal}

\end{document}